\documentclass[12pt,a4paper,aps,preprint,superscriptaddress,nofootinbib]{revtex4-1}
\usepackage{graphicx,color,verbatim,epsfig,subfigure}
\usepackage{slashed}
\usepackage{latexsym,amsfonts,amssymb,amsmath,makeidx,mathrsfs,multirow,bm,float}
\usepackage[makeroom]{cancel}
\usepackage[utf8]{inputenc}

\usepackage[colorlinks=true, pdfstartview=FitV, linkcolor=blue, citecolor=blue, urlcolor=blue]{hyperref}
\allowdisplaybreaks[4]

\def\beqn{\begin{eqnarray}}
\def\eeqn{\end{eqnarray}}
\def\beqs{\begin{subequations}}
\def\eeqs{\end{subequations}}
\def\beq{\begin{equation}}
\def\eeq{\end{equation}}
\def\ba{\begin{array}}
\def\ea{\end{array}}

\def\non{\nonumber\\}

\def\hf{\frac{1}{2}}
\def\[{\left[}
\def\]{\right]}
\def\({\left(}
\def\){\right)}

\def\TeV{\rm TeV}
\def\GeV{\rm GeV}

\def\gU{\rm U}
\def\gSU{\rm SU}

\def\mL{\mathcal{L}}

\def\mO{\mathcal{O}}

\begin{document}

\title{The future probe of the light Higgs boson pair production}

\author{Ning Chen}
\email{chenning$\_$symmetry@nankai.edu.cn}
\affiliation{
School of Physics, Nankai University, Tianjin 300071, China
}
\author{Tong Li}
\email{litong@nankai.edu.cn}
\affiliation{
School of Physics, Nankai University, Tianjin 300071, China
}
\author{Wei Su}
\email{wei.su@adelaide.edu.au}
\affiliation{ARC Centre of Excellence for Dark Matter Particle Physics and CSSM,
Department of Physics, University of Adelaide, SA 5005, Australia}
\author{Yongcheng Wu}
\email{ycwu@physics.carleton.ca}
\affiliation{Ottawa-Carleton Institute for Physics, \\
Carleton University, 1125 Colonel By Drive, Ottawa, Ontario K1S 5B6, Canada }

\date{\today}

\begin{abstract}
In this work we study the light Higgs scenario in the framework of two Higgs doublet model (2HDM). In this case the heavier CP-even Higgs boson $H$ in 2HDM is the SM-like Higgs with 125 GeV mass and a CP-even Higgs boson $h$ lighter than 125 GeV exhibits in the spectrum.
We find that this scenario exists in the alignment
limit of $\sin(\beta-\alpha)\simeq 0$ and is still allowed by the global fit of the 125 GeV Higgs and the direct searches for $H\to hh$ and $h\to \gamma\gamma$.
The case of $M_h<M_H/2$ is highly constrained as the exotic decay mode $H\to hh$ is kinematically allowed.
We focus on the $M_h>M_H/2$ case and simulate the pair production of the light Higgs boson at the LHC.
The pair production is sensitive to the soft $\mathbb{Z}_2$ symmetry breaking term $m_{12}$ in the 2HDM potential through the Higgs self-couplings.
We find that the future high-luminosity LHC can discover the $b\bar{b}\gamma\gamma$ signal in the region of $\tan\beta\lesssim 1$ and $0<m_{12}<100$ GeV allowed by the theoretical constraints.
\end{abstract}

\maketitle

%###################################################################
%%%%%%%%%%%%%%%%%%%%%%%%%%%%%%%%
\section{Introduction}
\label{section:intro}
%%%%%%%%%%%%%%%%%%%%%%%%%%%%%%%%

In new physics beyond the Standard Model (BSM), the setup with two Higgs doublets in the scalar sector is quite general.
Most studies of the 2HDM~\cite{Branco:2011iw} focus on the scenario where the SM-like Higgs boson with mass of $125\,\GeV$ is the lightest scalar $h$, while two additional neutral Higgs bosons $H$ and $A$ and two charged Higgs bosons $H^\pm$ are generally heavier.
The search for such heavy extra Higgs bosons requires the energy upgrade for the Large Hadron Collider (LHC) and even the future $100\,\TeV$ $pp$ collider~\cite{Craig:2016ygr,Kling:2018xud,Li:2020hao}.
In contrast, the situation with the heavy $H$ being the $125\,\GeV$ Higgs boson discovered at the LHC has been discussed in Refs.~\cite{Bernon:2014nxa, Bernon:2015wef,Cacciapaglia:2016tlr}.
In this case there exhibits a light Higgs boson $h$ escaping the LEP bound with the absent $hZZ$ coupling.
This scenario was also studied in the minimal Supersymmetric Standard Model (MSSM)~\cite{Christensen:2012ei,Christensen:2012si,Ke:2012zq} but suffers from strong flavor constraint as the mass of the charged Higgs bosons is close to 125 GeV~\cite{Han:2013mga}.
In the 2HDM, such constraint is weakened by the presence of heavy $H^\pm$ and $A$ bosons~\cite{Kling:2020hmi,Su:2019dsf}.

Recently, the CMS collaboration performed the direct search for a light Higgs boson in the mass range between
$70$ and $110$ GeV followed by the decay into diphoton~\cite{Sirunyan:2018aui}.
They reported a local (global) significance of 2.8 (1.3) standard deviations for a mass of $95.3$ GeV.
ATLAS also searched for low-mass diphoton resonances in the range of 65$-$110 GeV~\cite{ATLAS:2018xad}.
They found no significant excess with respect to the SM expectation. There also exist LHC searches for the exotic SM Higgs decay mode $H\to hh$~\cite{Sirunyan:2018pzn,Sirunyan:2018mot,Sirunyan:2019gou,Aaboud:2018gmx,Aaboud:2018esj} but no excess was observed~\cite{Gu:2017ckc,Chen:2018shg,Chen:2019pkq}.
Nevertheless, as the BSM new physics such as the 2HDM is compatible with the observed SM-like Higgs boson and can provide a new degree of freedom below 125 GeV, it is important to consider other search mode than the above single production channel $h\to \gamma\gamma$ or the SM-like Higgs exotic decays.

The Higgs boson pair production is usually viewed as the precision measurement of the Higgs self-coupling from the scalar potential.
The Higgs pair is produced through both the triangle diagrams governed by the Higgs self-coupling and the box diagram through a top quark loop~\cite{Plehn:1996wb,Shao:2013bz,Hespel:2014sla}.
The box diagram is determined by double Yukawa couplings with respect to one Yukawa coupling in the single production.
Moreover, the triangle processes are mediated by $H$ and $h$ via the $s$-channel.
Two classes of diagrams may also induce an interference with each other~\cite{Chen:2014xra,Carena:2018vpt}.
As a result, the Higgs boson pair process may exhibit an enhancement of the production cross section in some parameter space.
We thus propose the search potential for the light Higgs through the pair production channel at the LHC 14 TeV run.
We will first review the light Higgs scenario in the 2HDM and discuss the constraints from the SM-like Higgs exotic decay, the 125 GeV Higgs global fit and the direct LHC searches. The search potential of the light Higgs pair at future LHC upgrade is given by analyzing the signal and the SM backgrounds.

The rest of this paper is organized as follows.
In Sec.~\ref{section:2HDM}, we briefly review the setup of the 2HDM, where the SM-like Higgs boson with mass of $125\,\GeV$ is assumed to be a heavier CP-even one in the spectrum.
The relevant constraints are discussed in Sec.~\ref{section:constraints} and we show the allowed parameter space.
In Sec.~\ref{section:pair}, we calculate the total cross section of the light Higgs pair production and analyze the signal and background channels for some benchmark points passing the constraints.
The integrated luminosity needed at the LHC 14 TeV is given for the discovery of our benchmarks.
Finally, in Sec.~\ref{section:conclusion}, we summarize our conclusions.

%###################################################################
%%%%%%%%%%%%%%%%%%%%%%%%%%%%%%%%
\section{The CP-conserving two-Higgs-doublet model}
\label{section:2HDM}
%%%%%%%%%%%%%%%%%%%%%%%%%%%%%%%%

%%%%%%%%%%%%%%%%%%%%%%%%%%%%%%%%
\subsection{The 2HDM potential}
%%%%%%%%%%%%%%%%%%%%%%%%%%%%%%%%

The CP-conserving (CPC) 2HDM has two Higgs doublets $\Phi_{1\,,2}\in 2_{+1}$ under the electroweak gauge symmetry of $\gSU(2)_L\times \gU(1)_Y$.
The 2HDM potential at the tree level is given as follows
\beqn
V_0(\Phi_1\,,\Phi_2)&=&m_{11}^2|\Phi_1|^2+m_{22}^2|\Phi_2|^2- m_{12}^2 ( \Phi_1^\dag\Phi_2+H.c.)+\hf\lambda_1 |\Phi_{1}|^{4} +\hf\lambda_2|\Phi_{2}|^{4}\non
&+&\lambda_3|\Phi_1|^2 |\Phi_2|^2+\lambda_4 |\Phi_1^\dag \Phi_2|^2+\hf \lambda_5 \Big[ (\Phi_1^\dag\Phi_2)^2 +H.c.\Big]\,,
\eeqn
where all parameters are real for the CPC case.
The $\Phi_1^\dagger \Phi_2$ term with dimensional parameter $m_{12}$ softly breaks the global $\mathbb{Z}_2$ symmetry and leads to the spontaneous CP violation source~\cite{Davidson:2005cw,Gunion:2005ja,Grzadkowski:2013rza,Inoue:2014nva,Chen:2020soj} in the scalar sector.
After the electroweak symmetry breaking (EWSB), the mass spectrum of the 2HDM contains five physical Higgs bosons: $(h\,,H\,,A\,,H^\pm)$.
Two CP-even Higgs bosons are diagonalized by the mixing angle of $\alpha$, and the ratio of two Higgs doublet VEVs are parametrized by $\tan\beta\equiv v_2/v_1$.
Together, the physical mass inputs and mixing angles are related to the quartic self couplings in the generic basis by
\beqs\label{eqs:lambda_physical}
\beqn
\lambda_1&=& \frac{ 1 }{ v^2\, \cos^2\beta } (M_h^2 \sin^2\alpha + M_H^2 \cos^2\alpha - m_{12}^2 \tan \beta )\;, \\
\lambda_2&=& \frac{ 1 }{ v^2\, \sin^2\beta  } ( M_h^2 \cos^2\alpha + M_H^2 \sin^2\alpha - m_{12}^2 /\tan\beta )\;, \\
\lambda_3&=& \frac{1}{v^2} \Big[ \frac{ ( M_H^2 - M_h^2 ) \sin\alpha \cos\alpha }{ \sin\beta \cos\beta} + 2 M_{H^\pm}^2 - \frac{ m_{12}^2 }{ \sin\beta  \cos\beta } \Big]\;, \\
\lambda_4&=& \frac{1}{v^2} ( M_A^2 - 2 M_{H^\pm}^2 + \frac{ m_{12}^2 }{ \sin\beta \cos\beta} )\;,\\
\lambda_5&=&  \frac{1}{v^2}  ( \frac{ m_{12}^2 }{ \sin\beta \cos\beta}  - M_A^2 )\;.
\eeqn
\eeqs
In our discussions below, we shall assume that $M_H=125\,\GeV$ and $M_h < M_H < (M_A\,, M_{H^\pm})$.
In contrast to the usual case that the parameter choice of $\cos(\beta -\alpha)\to 0$ is favored where the light CP-even Higgs boson $h$ is assumed to be $125\,\GeV$, one can expect the different alignment limit of $\sin(\beta-\alpha)=0$ where $M_H=125\,\GeV$.
Correspondingly, we have
\beqs
\begin{eqnarray}
\lambda_1&=&{M_h^2\over v^2}-{  m_{12}^2-M_H^2 \sin\beta \cos\beta \over v^2 \cos^2\beta } \tan\beta +{M_h^2-M_H^2\over v^2} ( \tan^2\beta -1)\, , \\
\lambda_2&=&{M_h^2\over v^2}-{m_{12}^2-M_H^2 \sin\beta \cos\beta\over v^2 \tan\beta \sin^2\beta }+{M_h^2-M_H^2\over v^2} (\frac{1}{\tan^2 \beta }-1)\, ,\\
\lambda_3&=&{-M_h^2+2M_{H^\pm}^2\over v^2}-{m_{12}^2-M_H^2 \sin\beta \cos\beta\over v^2 \sin\beta \cos\beta}\, ,\\
\lambda_4&=& {M_A^2-2M_{H^\pm}^2+M_{H}^2\over v^2}+{m_{12}^2-M_H^2 \sin\beta \cos\beta\over v^2 \sin\beta \cos\beta}\,,\\
\lambda_5&=& {M_H^2-M_A^2\over v^2}+{m_{12}^2-M_H^2 \sin\beta \cos\beta\over v^2 \sin\beta  \cos\beta}\,.
\end{eqnarray}
\eeqs
Besides of the fixed inputs of $M_H=125$ GeV and $\sin(\beta-\alpha)=0$, we perform a scan over the rest of 2HDM parameter space by using 2HDMC~\cite{Eriksson:2009ws,Harlander:2013qxa}
\begin{eqnarray}
0.1&<&\tan\beta<60, \ 10 \ {\rm GeV}<M_h<120 \ {\rm GeV}, \nonumber \\
200 \ {\rm GeV}&<&M_A=M_{H^\pm}<2000 \ {\rm GeV}, \ 0<m_{12}^2<( 100 \ {\rm GeV})^2 \; ,
\label{scanspace}
\end{eqnarray}
under the consideration of vacuum stability and unitarity.
The requirement of electroweak precision measurements is satisfied by the mass degeneracy $M_A=M_{H^\pm}$.

%%%%%%%%%%%%%%%%%%%%%%%%%%%%%%%%
\subsection{The couplings of Higgs bosons in the 2HDM}
%%%%%%%%%%%%%%%%%%%%%%%%%%%%%%%%

In the general 2HDM, there could be tree-level flavor-changing neutral currents (FCNC), which are well-known constraints on such model.
To alleviate the tree-level FCNC process constraints, the SM fermions of a given representation are usually assigned to a single Higgs doublet.
We focus on the Type-I and Type-II Yukawa couplings of
\beqn
\mL&\supset& \sum_{h_i=h\,,H}  - \frac{m_f}{v} \left( \xi_i^f \bar f f   h_i + \xi_A^f \bar f i \gamma_5 f A \right) \,,
\eeqn
with
\beqs\label{eqs:Higgs_coup}
\beqn
\textrm{Type-I}&:& \xi_h^f= \sin(\beta-\alpha) + \frac{ \cos(\beta-\alpha) }{\tan \beta} \,,\qquad \xi_{H}^f=  \cos(\beta-\alpha) - \frac{ \sin(\beta-\alpha) }{ \tan\beta}\non
&& \xi_A^u= \frac{1}{ \tan\beta}\,, \qquad \xi_A^{d\,, \ell}= - \frac{1}{ \tan\beta}\,,\\
\textrm{Type-II}&:&\xi_h^u= \sin(\beta-\alpha) + \frac{ \cos(\beta-\alpha) }{ \tan\beta}\,,\qquad \xi_h^{d\,,\ell}= \sin(\beta-\alpha) - \cos(\beta-\alpha)\, \tan\beta \,,\non
&&\xi_H^u= \cos(\beta-\alpha) - \frac{ \sin(\beta-\alpha) }{ \tan\beta}\,,\qquad \xi_H^{d\,,\ell}=  \cos(\beta-\alpha) + \sin( \beta-\alpha) \,\tan\beta\,,\non
&& \xi_A^u= \frac{1}{ \tan\beta}\,,\qquad \xi_A^{d\,,\ell} = \tan\beta\,.
\eeqn
\eeqs
Under the limit of $\sin(\beta - \alpha)=0$, the normalized Yukawa couplings are reduced to
\beqs\label{eqs:Higgs_coup_align}
\beqn
\textrm{Type-I}&:& \xi_h^f=  \frac{ 1 }{\tan \beta} \,,\qquad \xi_{H}^f=  1 \non
&& \xi_A^u= \frac{1}{ \tan\beta}\,, \qquad \xi_A^{d\,, \ell}= - \frac{1}{ \tan\beta}\,,\\
\textrm{Type-II}&:&\xi_h^u=  \frac{ 1 }{ \tan\beta}\,,\qquad \xi_h^{d\,,\ell}=  -  \tan\beta \,,\non
&&\xi_H^u= 1 \,,\qquad \xi_H^{d\,,\ell}=  1 \,,\non
&& \xi_A^u= \frac{1}{ \tan\beta}\,,\qquad \xi_A^{d\,,\ell} = \tan\beta\,.
\eeqn
\eeqs
Besides, two CP-even Higgs bosons couple to the gauge bosons such that
\beqn
\mL&\supset& \sum_{h_i=h\,,H}  a_i \left( 2  \frac{m_W^2}{v} W_\mu^+ W^{-\,\mu} +  \frac{m_Z^2}{v} Z_\mu Z^\mu \right)  h_i \non
&+& ( g_{hhZZ} h^2  + g_{HHZZ} H^2 ) Z^\mu Z_\mu \,,
\eeqn
with
\beqn
&&a_h= \sin(\beta-\alpha)\,,\quad a_H = \cos(\beta-\alpha) \,,\quad g_{hhZZ} = g_{HHZZ} = \frac{m_Z^2}{2 v^2} \,.
\eeqn

%%%%%%%%%%%%%%%%%%%%%%%%%%%%%%%%
\subsection{The cubic scalar self couplings of 2HDM}
%%%%%%%%%%%%%%%%%%%%%%%%%%%%%%%%

\begin{figure}[htb]
\centering
\includegraphics[width=0.45\textwidth]{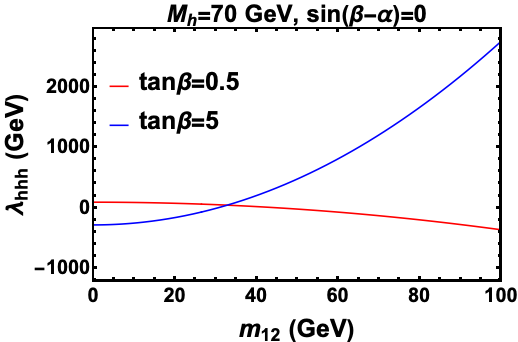}
\includegraphics[width=0.45\textwidth]{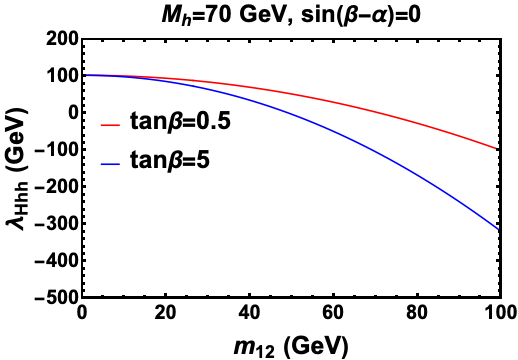}
\caption{
The cubic scalar self couplings of $\lambda_{hhh}$ (left) and $\lambda_{Hhh}$ (right) versus the soft $\mathbb{Z}_2$ breaking mass term $m_{12}$.
We fix $M_h=70\,\GeV$ and take $\tan\beta=0.5$ (red) and $\tan\beta=5.0$ (blue) in both plots.
}
\label{fig:lambdas_lighth}
\end{figure}

In the physical basis, we list the cubic scalar self couplings below~\cite{Kanemura:2004mg,Hespel:2014sla}
\beqs
\beqn
\lambda_{hhh}&=&\frac{3!}{v} \Big[  \hf M_h^2 \Big( \sin(\beta-\alpha) + 2  \sin(\beta-\alpha) \cos^2(\beta-\alpha) + \frac{2 \cos^3(\beta-\alpha) }{ \tan(2\beta) }  \Big) \non
&-& \frac{m_{12}^2}{ \sin\beta \cos\beta} \cdot  \Big( \sin(\beta-\alpha) + \frac{ \cos( \beta-\alpha) }{ \tan(2\beta)  } \Big) \cos^2(\beta-\alpha)   \Big] \,,\\
\lambda_{Hhh}&=& \frac{ \cos(\beta-\alpha) }{v} \Big[ ( M_H^2 + 2 M_h^2  ) \cdot\Big( -1  + 2 \cos^2(\beta-\alpha) - \frac{ 2  \sin(\beta-\alpha) \cos(\beta-\alpha)  }{ \tan(2\beta) } \Big)  \non
&+& \frac{2 m_{12}^2 }{ \sin\beta \cos\beta} \cdot \Big( 2 + \frac{ 3 \sin(\beta-\alpha) \cos(\beta-\alpha)  }{\tan(2\beta) } - 3 \cos^2(\beta-\alpha)  \Big)  \Big] \,,\\
\lambda_{hHH}&=& \frac{ \sin(\alpha-\beta) }{ v} \Big[  (M_h^2 + 2 M_H^2 ) \frac{ \sin(2\alpha) }{ \sin(2\beta) } \non
&-& \frac{2 m_{12}^2}{ \sin\beta \cos\beta } \Big( 2 - \frac{3 \sin(\beta-\alpha)  \cos(\beta-\alpha) }{ \tan( 2\beta)  } - 3 \sin^2(\beta-\alpha) \Big) \Big]\,,
\eeqn
\eeqs
with the Feynman rules being $-i\lambda_{H_iH_jH_k}$.
Under the limit of $\sin(\beta-\alpha) = 0$, they are reduced to
\beqs
\beqn
\lambda_{hhh}&\to&\frac{3!}{v \tan(2\beta) } (  M_h^2  - \frac{m_{12}^2}{ \sin\beta \cos\beta} ) \,,\\
\lambda_{Hhh}&\to& \frac{1 }{v} ( M_H^2 + 2 M_h^2   - \frac{2 m_{12}^2 }{  \sin\beta \cos\beta  }  ) \,.
%,\\
%\lambda_{hHH}&\to& 0 \,.
\eeqn
\eeqs
In Fig.~\ref{fig:lambdas_lighth}, we display the cubic scalar self couplings of $\lambda_{hhh}$ and $\lambda_{Hhh}$ versus $m_{12}$, with the fixed inputs of $\tan\beta=0.5$ and $\tan\beta = 5$, respectively.
It turns out that the variation of $\lambda_{hhh}$ is moderate with the small input parameter of $\tan\beta =0.5$, as compared with the large input parameter of $\tan\beta =5$.
The absolute value of the cubic scalar self coupling of $\lambda_{Hhh}$ increase along with larger $m_{12}\gtrsim 40\sim 60$ GeV.

%###################################################################
%%%%%%%%%%%%%%%%%%%%%%%%%%%%%%%%
\section{The constraints to the light Higgs boson scenario}
\label{section:constraints}
%%%%%%%%%%%%%%%%%%%%%%%%%%%%%%%%

%%%%%%%%%%%%%%%%%%%%%%%%%%%%%%%%
\subsection{The exotic decay of the SM-like Higgs boson}
%%%%%%%%%%%%%%%%%%%%%%%%%%%%%%%%

The presence of a light Higgs boson leads to the exotic decay mode of $H\to hh$ for $M_H>2M_h$, with the corresponding on-shell partial decay width being
\beqn
\Gamma[H\to hh]&=& \frac{\lambda_{Hhh}^2}{32\pi M_H} \sqrt{1- 4 \kappa_h }\, ,
\eeqn
with $\kappa_h \equiv M_h^2/M_H^2$.
For the $2 M_h > M_H > M_h$ case, the heavy Higgs boson $H$ decays to an on-shell light Higgs boson $h$ and an off-shell one, followed by the latter coupling to SM fermion pairs.
The partial width of this three-body decay is given by~\cite{Djouadi:1995gv}
\beqn
\Gamma[H\to hh^* \to h \bar f f ]&=& \frac{ N_{c\,,f}}{ 32 \pi^3  M_H  } \lambda_{Hhh}^2 (\xi_{h}^f)^2 {m_f^2\over v^2}  \Big[  ( \kappa_h -1 ) \Big(2 - \hf \log \kappa_h \Big) \non
&+& \frac{ 1 - 5 \kappa_h }{ \sqrt{4 \kappa_h -1} } \Big( \tan^{-1}\Big( \frac{ 2 \kappa_h -1 }{ \sqrt{4 \kappa_h -1} }\Big) - \tan^{-1}\Big( \frac{ 1 }{ \sqrt{4 \kappa_h -1} }\Big) \Big)  \Big],
\eeqn
where $N_{c\,,f}=3 \ (1)$ for SM quarks (leptons). The off-shell decay partial widths are typically negligible.

\begin{figure}[htb]
\centering
\includegraphics[width=0.45\textwidth]{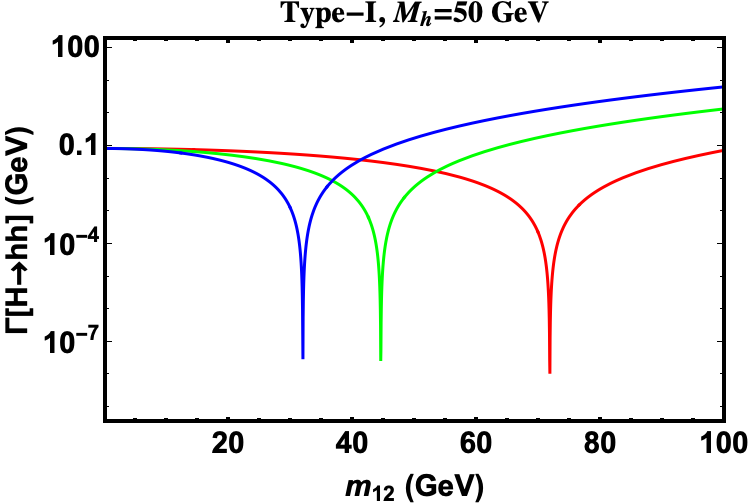}
\includegraphics[width=0.45\textwidth]{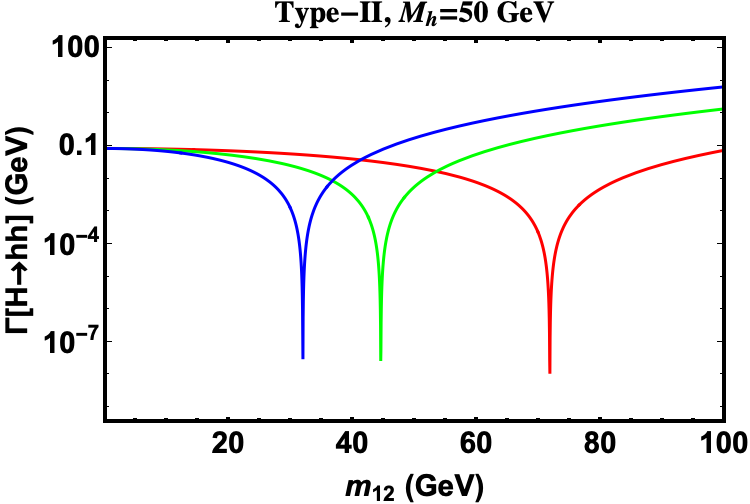}\\
\includegraphics[width=0.45\textwidth]{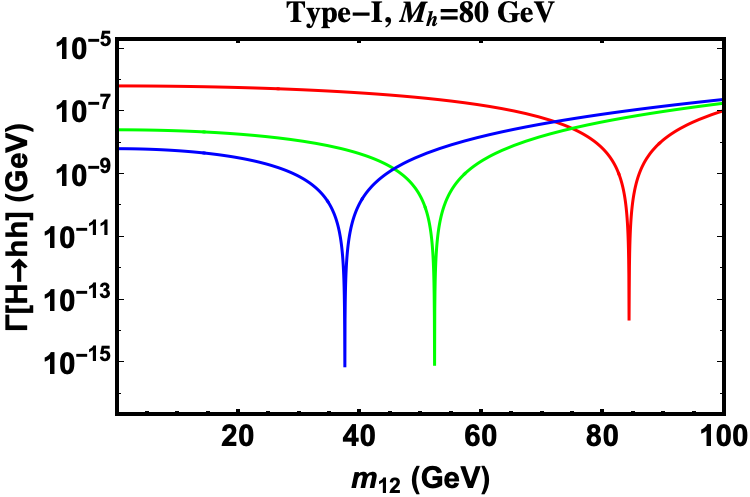}
\includegraphics[width=0.45\textwidth]{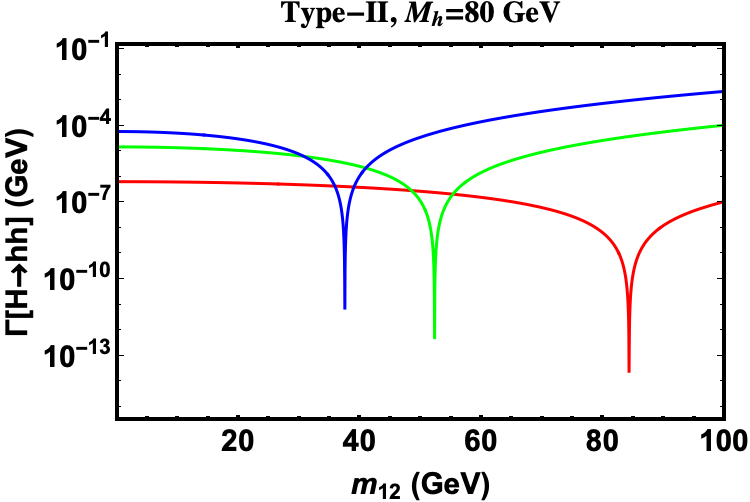}
\caption{
The partial decay widths of $\Gamma[H\to hh]$ versus $m_{12}$, with $M_h=50\,\GeV$ (upper panels) and $M_h=80\,\GeV$ (lower panels).
Three curves are shown for the $\tan\beta=1.0$ (red), $\tan\beta=5.0$ (green), and $\tan\beta=10.0$ (blue) cases, together with $\sin(\beta - \alpha)=0$.
}
\label{fig:Hhh_width}
\end{figure}

The direct measurements of the SM-like Higgs boson decay width was made by ATLAS in Ref.~\cite{Aad:2014aba}.
It was reported that the upper limits to the total width of the Higgs boson are $5.0\,\GeV$ from the $H\to \gamma\gamma$ channel, and $2.6\,\GeV$ from the $H\to ZZ^*\to 4\ell$ channel, respectively.
In Fig.~\ref{fig:Hhh_width}, the partial decay widths of $\Gamma[H\to hh]$ are shown for both on-shell decay mode (with $M_h=50\,\GeV$) and off-shell decay mode (with $M_h=80\,\GeV$).
For the on-shell decay modes with $M_h=50\,\GeV$, the partial decay width of $\Gamma[H \to h h ]$ can be as large as $\sim \mO (0.1)- \mO(10)\,\GeV$.
When the light Higgs boson mass is $M_h=80\,\GeV$, the off-shell partial decay width of $\Gamma[H \to h h ]$ are suppressed to $\sim \mO(10^{-9}) - \mO(10^{-4})\,\GeV$. Both decay modes exhibit a dip when $\lambda_{Hhh}\to 0$ with some value of $m_{12}$.

%%%%%%%%%%%%%%%%%%%%%%%%%%%%%%%%
\subsection{The LHC $125\,\GeV$ Higgs boson signal fit}
\label{subsec:LHC125fit}
%%%%%%%%%%%%%%%%%%%%%%%%%%%%%%%%

As the exotic decay $H\to hh$ changes the total width of the SM-like Higgs boson, the global signal fit to the heavy $H$ would lead to the constraint on the light Higgs scenario.
The global signal fit to the heavy $H$ as the $125\,\GeV$ Higgs boson signals at the LHC run-I and run-II is displayed below.
In Fig.~\ref{fig:H125fit_runIrunII_tbm12}, the global signal fit to the heavy $H$ as the $125\,\GeV$ Higgs boson is performed in the $(m_{12}\,, \tan\beta )$ plane, with $\sin(\beta - \alpha )=0$ assumed.
For the $M_h=50\,\GeV$ case, the allowed ranges of $m_{12}$ are highly correlated with $\tan\beta$, which correspond to the dips that were present in the upper panels in Fig.~\ref{fig:Hhh_width}.
For the $M_h=80\,\GeV$ case, the allowed ranges of $(m_{12}\,, \tan\beta)$ are significantly extended, with the suppressed off-shell decay mode of $H\to h h^*$.

\begin{figure}[htb]
\centering
\includegraphics[width=0.45\textwidth]{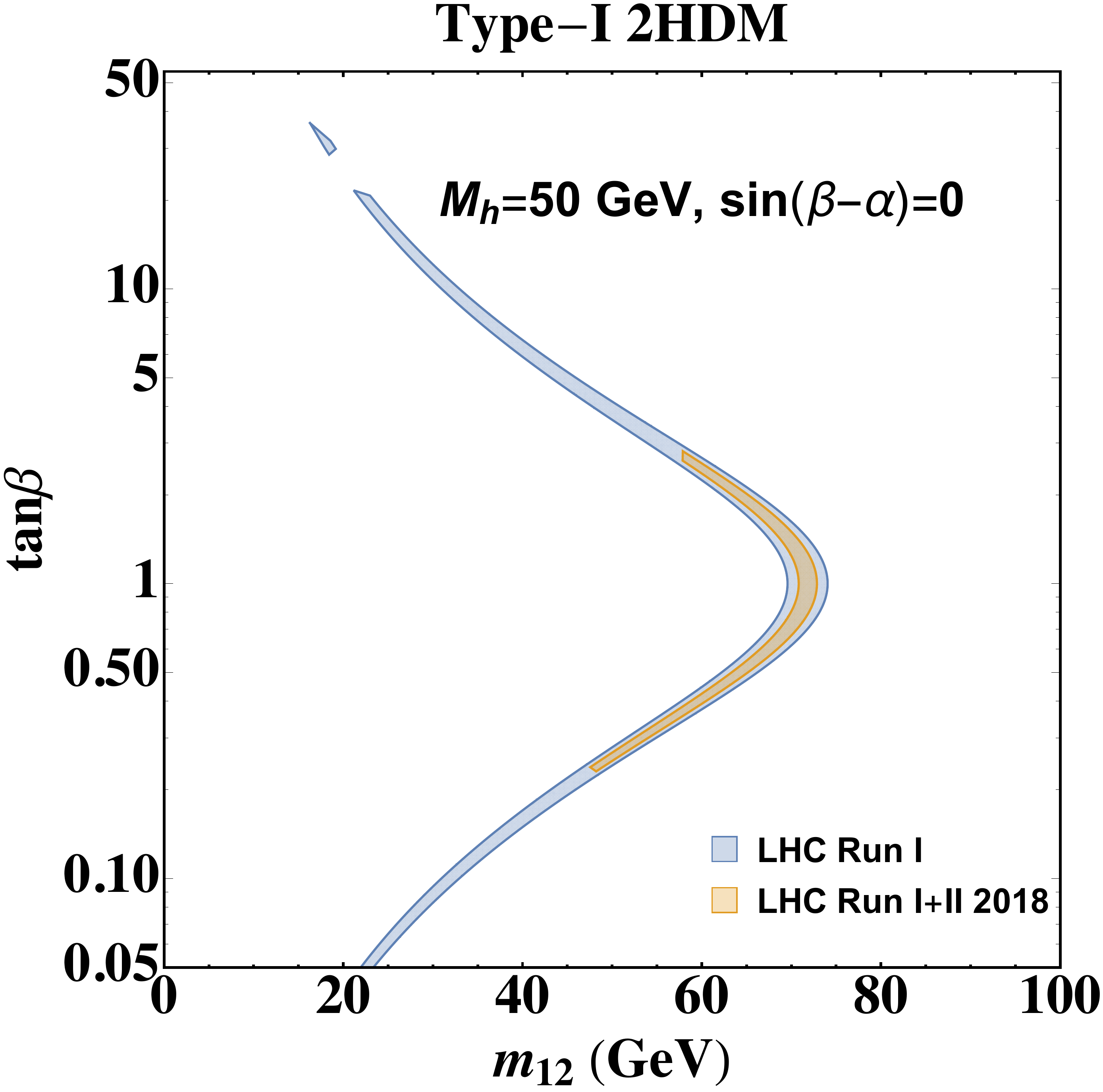}
\includegraphics[width=0.45\textwidth]{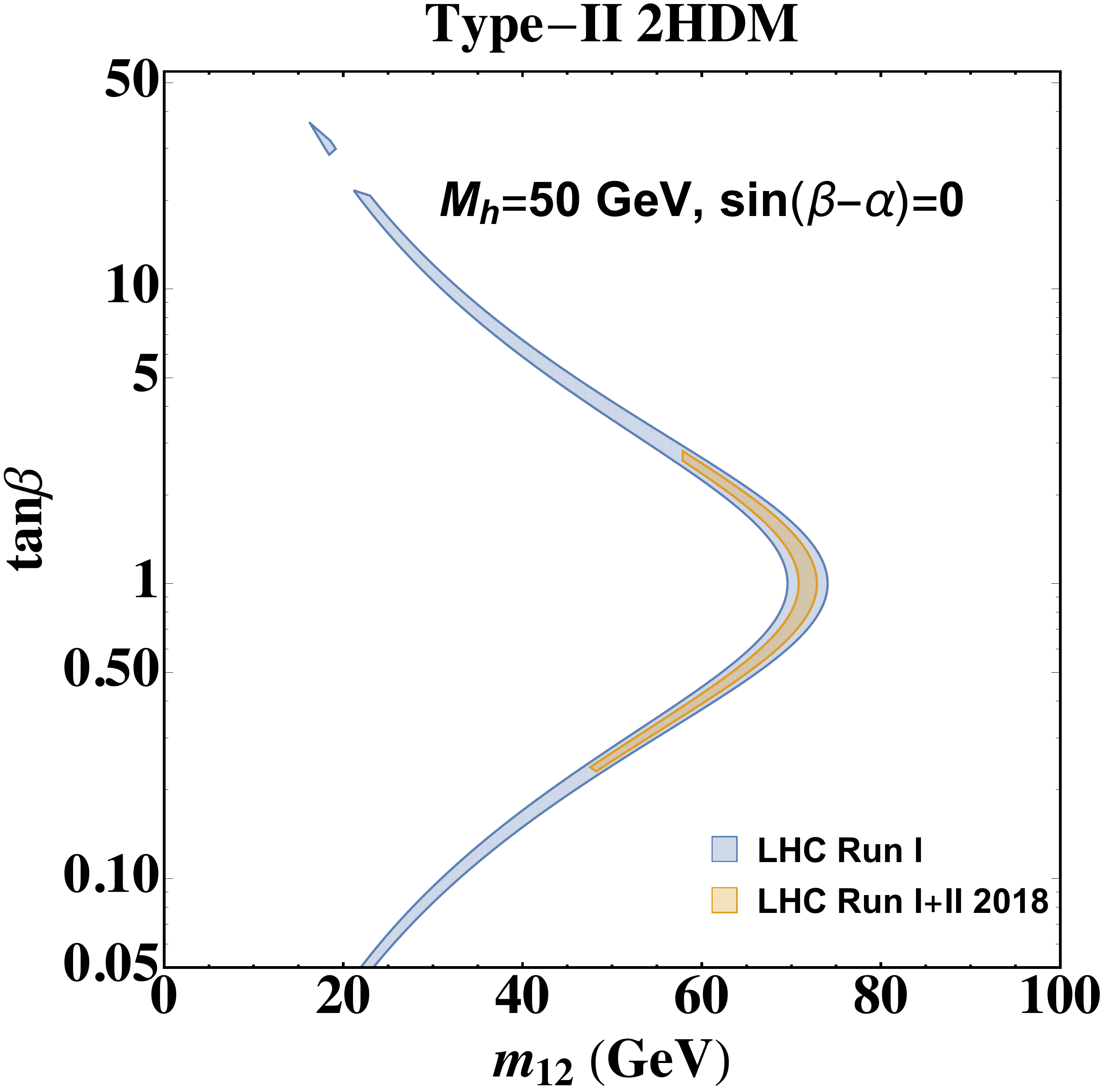}\\
\includegraphics[width=0.45\textwidth]{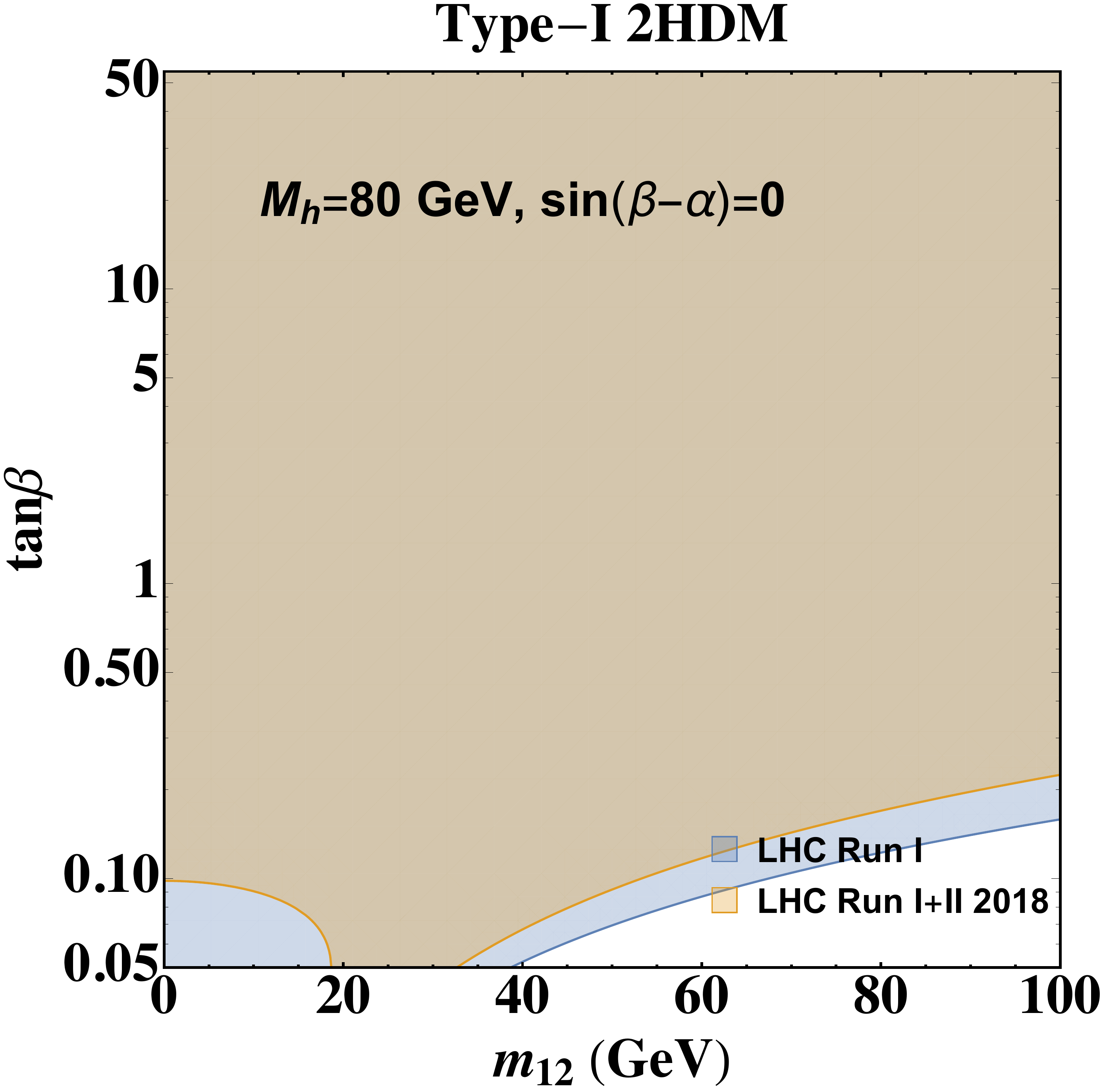}
\includegraphics[width=0.45\textwidth]{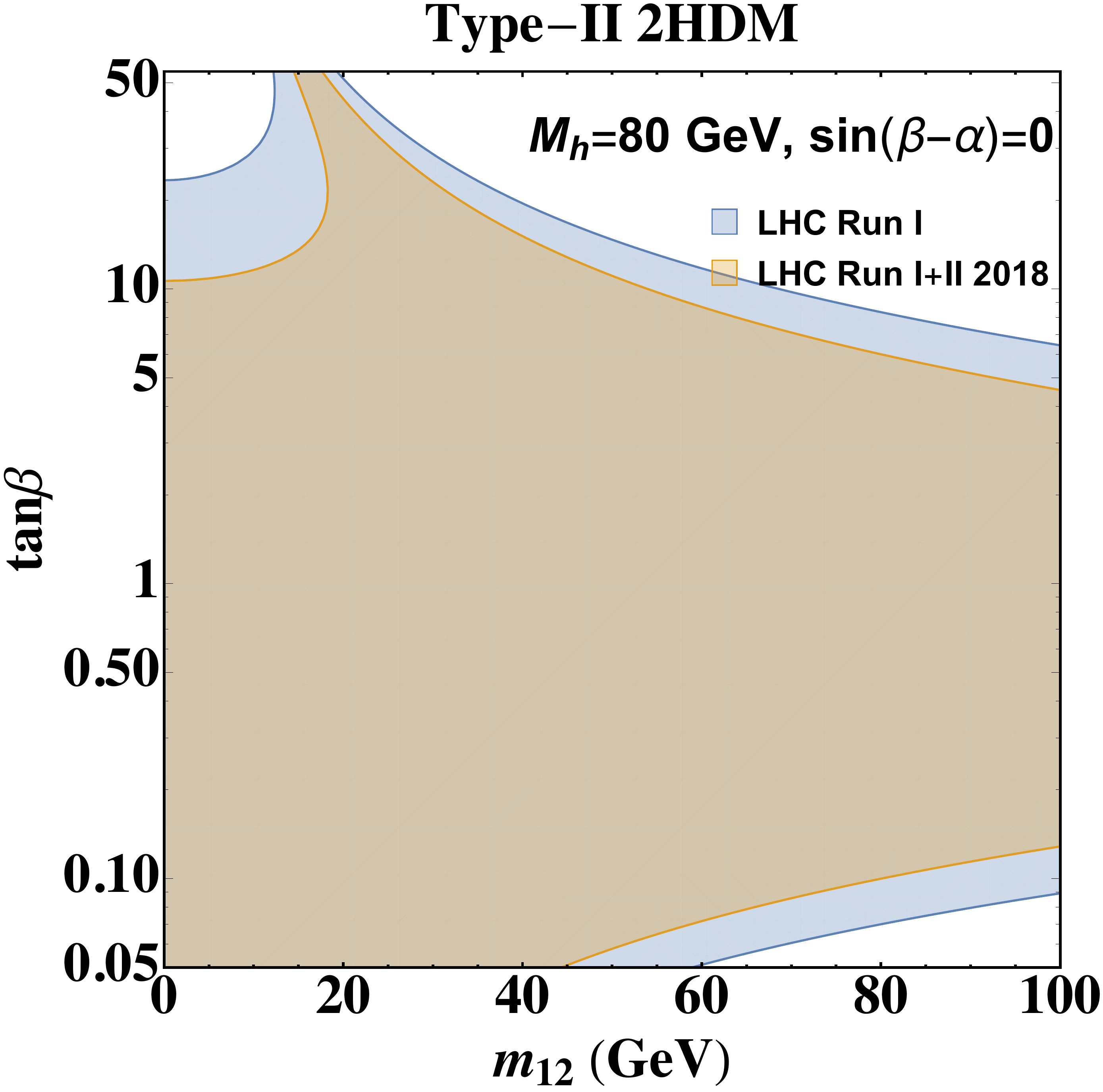}
\caption{
The global signal fit to the heavy $H$ as the $125\,\GeV$ Higgs boson at the LHC run-I and run-II, in the plane of $m_{12}$ versus $\tan\beta$ for $M_h=50$ GeV (up) or $M_h=80$ (bottom).
Both Type-I (left) and Type-II (right) cases are displayed.
We take $\sin(\beta-\alpha)=0$.
}
\label{fig:H125fit_runIrunII_tbm12}
\end{figure}

%%%%%%%%%%%%%%%%%%%%%%%%%%%%%%%%
\subsection{The direct LHC experimental searches for the light Higgs boson}
\label{subsec:LHClight}
%%%%%%%%%%%%%%%%%%%%%%%%%%%%%%%%

The current LHC experiments from run-I and run-II have carried out direct searches for the light Higgs bosons.
We take into account the direct LHC search constraints on the decay of 125 GeV Higgs $H\to hh$~\cite{Sirunyan:2018pzn,Sirunyan:2018mot,Sirunyan:2019gou,Aaboud:2018gmx,Aaboud:2018esj} and the light Higgs decay into diphoton $h\to \gamma\gamma$~\cite{Sirunyan:2018aui,ATLAS:2018xad}. For $M_h<M_H/2$ case, we further require $|\lambda_{Hhh}|<1$ GeV to suppress the exotic decay $H\to hh$ and satisfy the global fit implications.

\begin{figure}[htb]
\begin{center}
\includegraphics[width=0.45\textwidth]{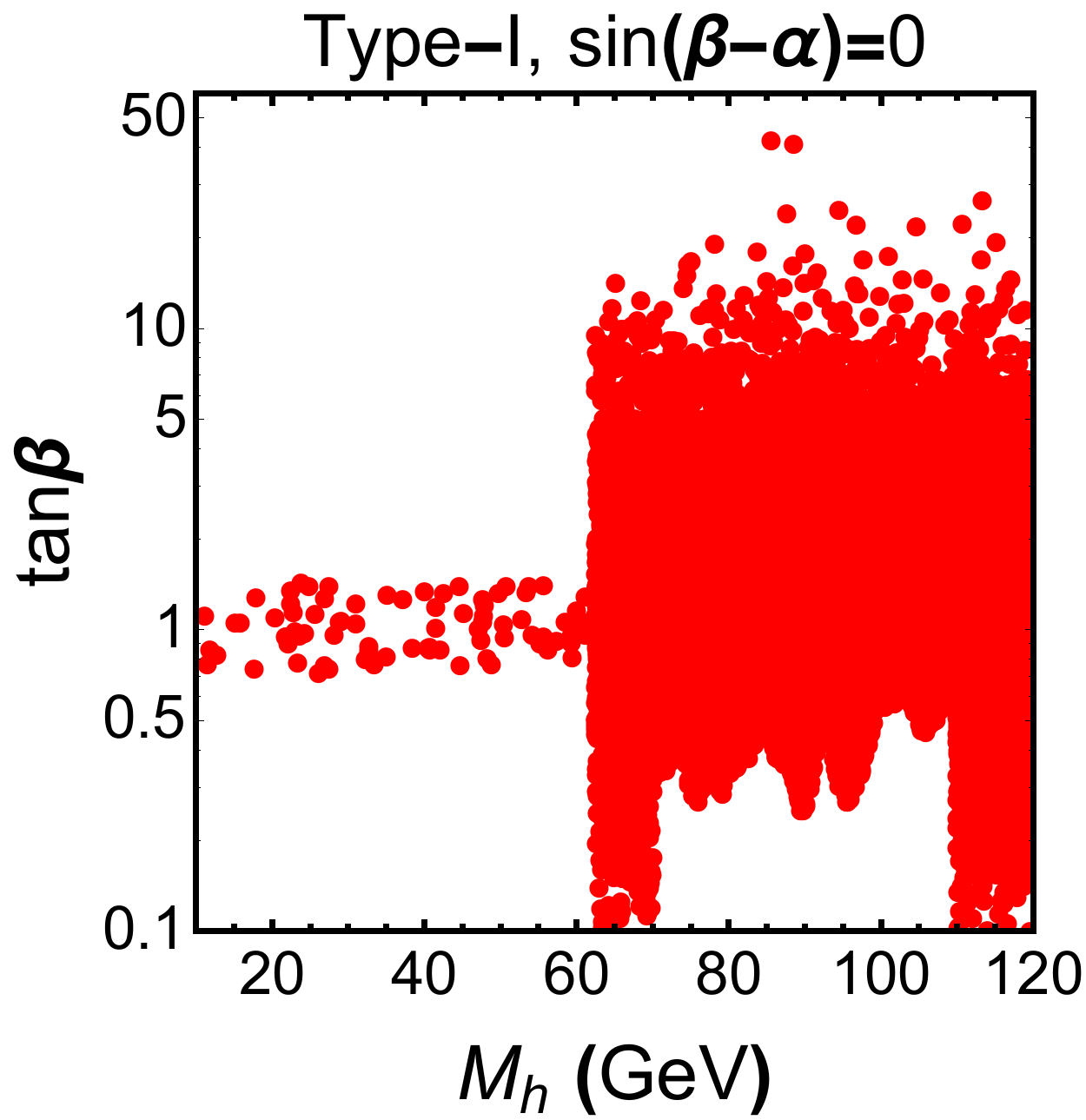}
\includegraphics[width=0.45\textwidth]{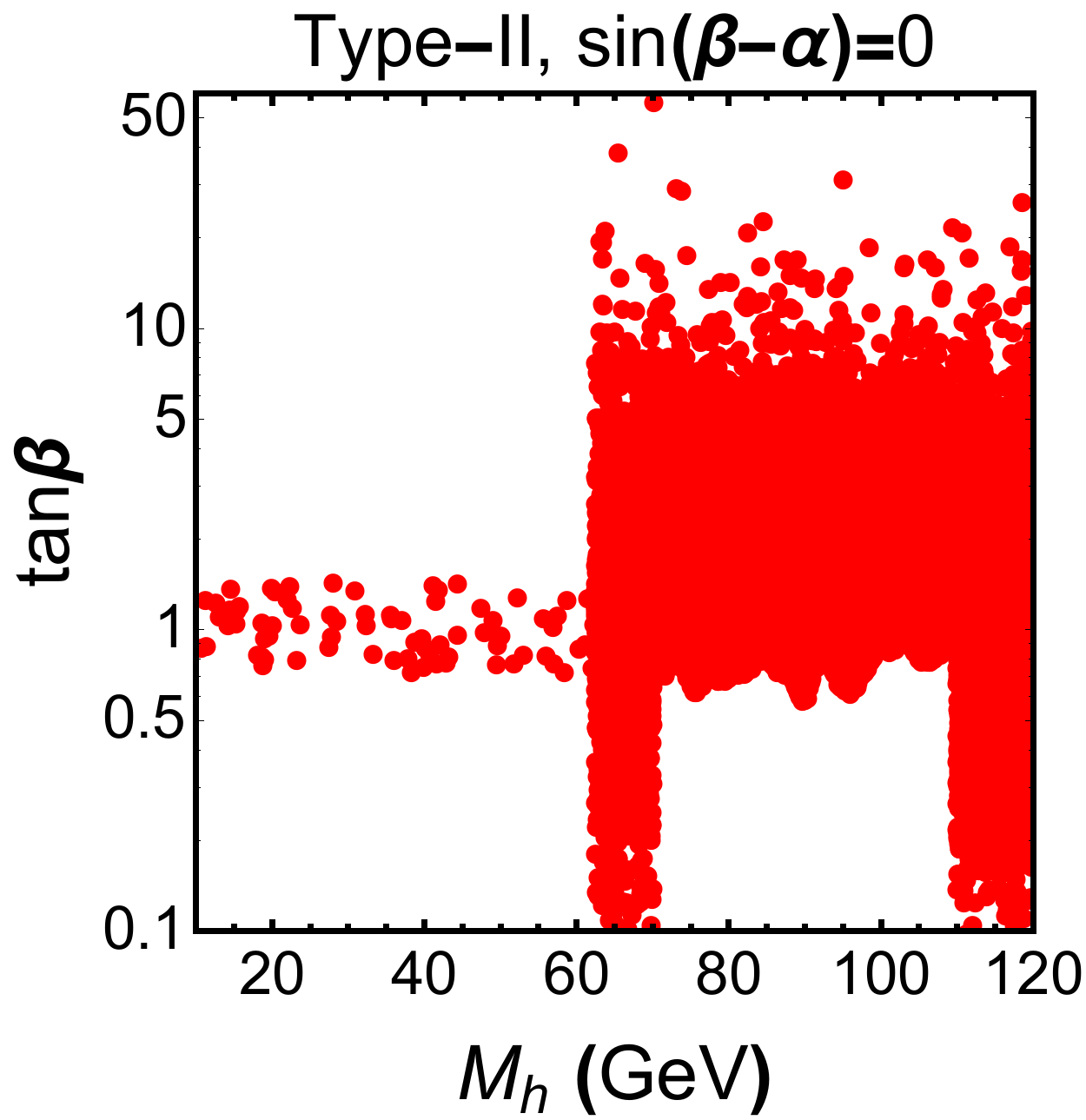}
\end{center}
\caption{The allowed parameter space in the plane of $M_h$ versus $\tan\beta$ after considering theoretical constraints, the direct LHC search constraints and $|\lambda_{Hhh}|<1$ GeV for $M_h<M_H/2$ case are imposed to satisfy the global fit.}
\label{fig:LHCconstraints}
\end{figure}

In Fig.~\ref{fig:LHCconstraints} we display the limits to the light CP-even Higgs boson in the $(M_h\,, \tan\beta)$ plane satisfying the above constraints.
In the case of $M_H>2M_h$, $\lambda_{Hhh}$ is highly constrained by the direct search for $H\to hh$ and one can see that only a small region of $\tan\beta\simeq 1$ is survived.
For $M_H<2M_h$, except the two extremely kinematic limits $M_h\simeq M_H/2$ and $M_H$, small $\tan\beta$ region is excluded by the $h\to \gamma\gamma$ search as the coupling to the top quark is enhanced.

%%%%%%%%%%%%%%%%%%%%%%%%%%%%%%%%
\subsection{The theoretical constraints with the light Higgs boson}
\label{subsec:theoretical}
%%%%%%%%%%%%%%%%%%%%%%%%%%%%%%%%

The theoretical constraints to the 2HDM potential include the perturbative unitarity and the stability bounds.
Roughly speaking, the perturbative unitarity requires that the scalar self-couplings in the 2HDM potential should not be too large to hit the Landau pole.
The more detailed constraint is usually obtained by evaluating the $S$-matrices for the scattering processes of the scalar fields in the 2HDM~\cite{Arhrib:2000is,Kanemura:2015ska}.
The stability constraints require a positive 2HDM potential for large values of Higgs fields along all field space directions.
Collectively, they read
\beqn\label{eq:2HDM_stability}
&& \lambda_{1\,,2}>0\,, \qquad \lambda_3 > - \sqrt{\lambda_1 \lambda_2  }\,,\qquad \lambda_3 + \lambda_4 - | \lambda_5 | > - \sqrt{ \lambda_1 \lambda_2  } \,.   
\eeqn
With the special limit of $\sin(\beta -\alpha)=0$, the requirements of $\lambda_{1\,,2}>0$ transform into the upper bound to the $m_{12}^2$ as
\beqn\label{eq:m12_upper}
&& m_{12}^2 \leq {\rm min}  ( M_h^2  + M_H^2 \tan^2\beta   \,, M_h^2 + \frac{M_H^2 }{ \tan^2\beta } ) \sin\beta \cos\beta \,,
\eeqn
with $M_h< M_H = 125\,\GeV$.

%###################################################################
%%%%%%%%%%%%%%%%%%%%%%%%%%%%%%%%
\section{The light Higgs pair production at the $pp$ collider}
\label{section:pair}
%%%%%%%%%%%%%%%%%%%%%%%%%%%%%%%%

\begin{figure}[!tbp]
\begin{center}
\includegraphics[width=\textwidth]{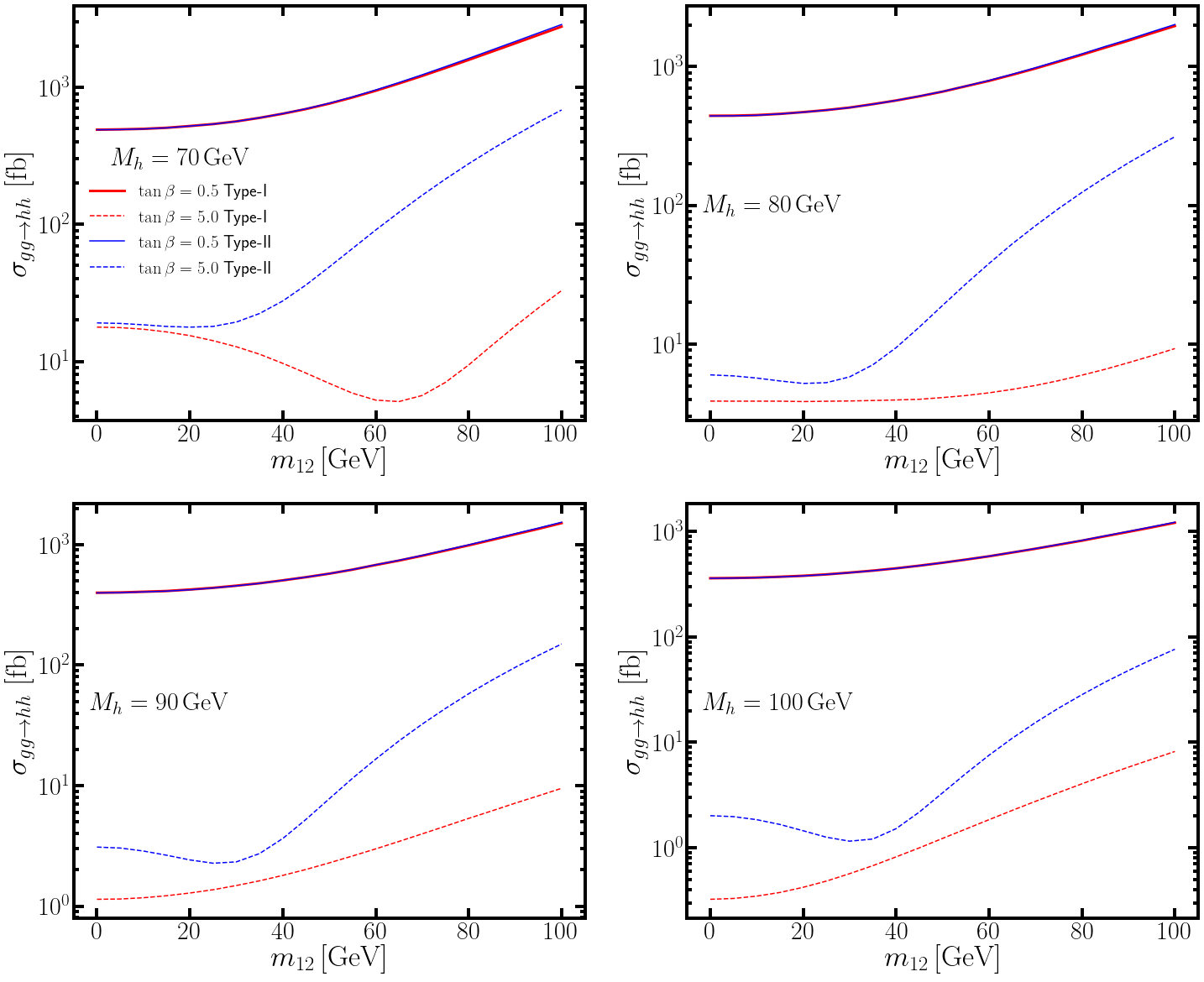}
\end{center}
\caption{
The total cross section of $\sigma_{gg\to hh}$ at the LHC $14\,\TeV$ run as a function of $m_{12}$ for $M_h>M_H/2$ case.
}
\label{fig:pphhxsec}
\end{figure}

\begin{figure}
\centering
\includegraphics[width=\textwidth]{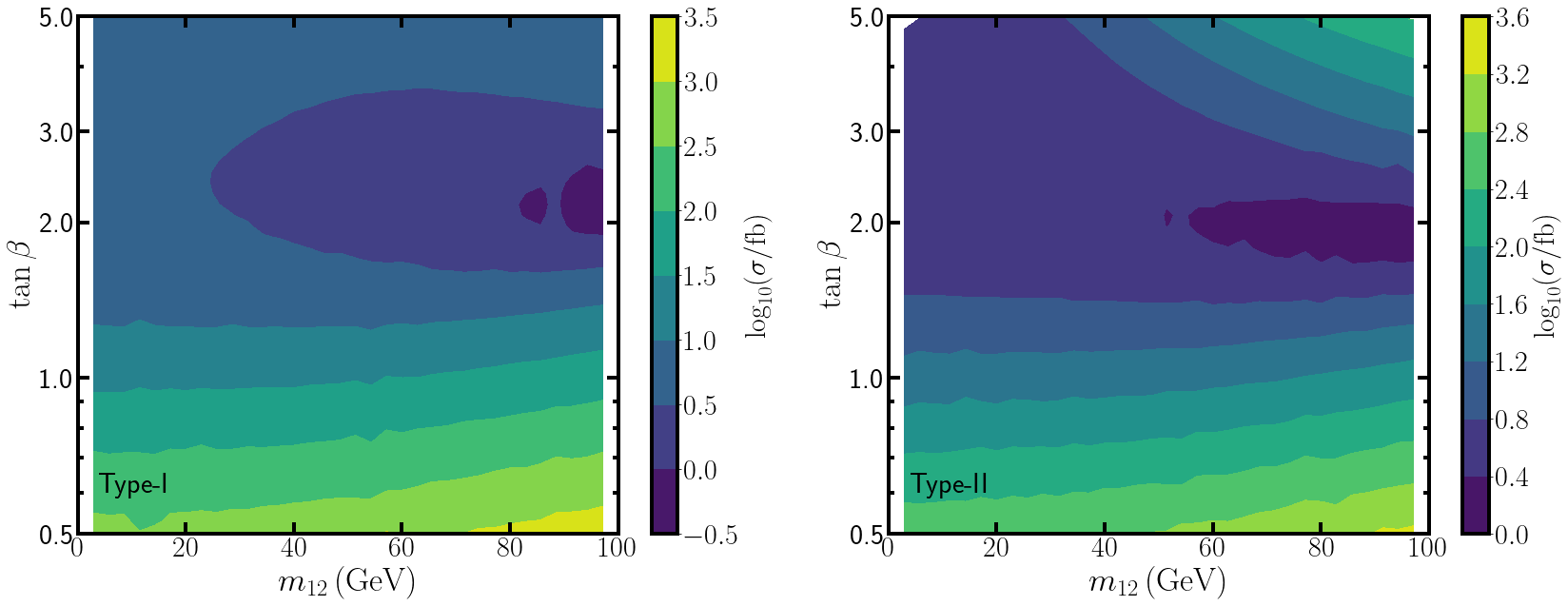}
\caption{The total cross section of $\sigma_{gg\to hh}$ at the LHC $14\,\TeV$ run in $( m_{12}\,, \tan\beta)$ plane for $M_h=80$ GeV case. Both Type-I (left) and Type-II (right) cases are displayed. We take $\sin(\beta-\alpha)=0$ and $M_{H^\pm}=M_A=800$ GeV.}
\label{fig:xsecMh80}
\end{figure}

The Higgs boson pairs are dominantly produced through two classes of diagrams: (1) the triangle diagram in which an $s-$channel Higgs mediates the two gluons transition to two Higgs
bosons, and (2) the box diagram in which the annihilation of two gluons through a top quark loop produces the Higgs boson pairs.
The triangle diagrams are determined by the soft $\mathbb{Z}_2$ symmetry breaking parameter $m_{12}$ through the cubic Higgs self couplings.
When $M_h<M_H/2$, in principle there are two triangle processes contributing to the pair production of two lighter Higgs bosons, i.e. $pp\to H, h^\ast \to hh$.
The SM-like Higgs boson $H$ is on-shell for the exotic decay $H\to hh$ in this case and as discussed above, the cubic coupling $\lambda_{Hhh}$ is required to be very small to be consistent with the global fit result. As a result, the surviving parameter region is limited and there would be only one intervening triangle process in this case.
We thus focus on the $M_h>M_H/2$ case in the following analysis of light Higgs boson pair production at the $pp$ collider.
For $M_h>M_H/2$, we have two triangle processes of $pp\to H^\ast, h^\ast\to hh$ with both off-shell intermediate Higgs bosons besides of the box diagram.
In Fig.~\ref{fig:pphhxsec} we show the total cross sections of $gg\to hh$ at the LHC $14\,\TeV$ run as a function of $m_{12}$ for fixed values of $\tan\beta=(0.5\,, 5.0)$ and $M_h=(70\,,80\,, 90\,, 100)\,\GeV$.
For the small $\tan\beta$ inputs, the $h$ couplings to the top quark are enhanced for both the Type-I and Type-II cases, and thus the box diagram is dominant.
As a result, the total cross section of $gg\to hh$ has weak dependence of $m_{12}$ and the results of Type-I and Type-II 2HDM are equivalent.
For large $\tan\beta$ inputs, the total cross sections strongly depend on $m_{12}$ but exhibit at least a few times smaller than that with small $\tan\beta$ for a fixed $m_{12}$.
Meanwhile, the cross sections of Type-II 2HDM are typically larger than those of Type-I as the $hdd$ coupling is enhanced by $\tan\beta$ in the Type-II 2HDM.
In Fig.~\ref{fig:xsecMh80}, we fix $M_h=80$ GeV and display the total cross section in the $(m_{12}\,, \tan\beta)$ plane.
One can see that cross sections are larger than 10 fb for $\tan\beta\lesssim 1.2$ and the maximal value occurs for large $m_{12}$ due to the enhancement of cubic self couplings.

As were previously suggested by the LHC analysis of the SM-like Higgs pair production~\cite{Aad:2019uzh}, among the final states of $b\bar{b}b\bar{b}$, $b\bar{b}\tau^+\tau^-$ and $b\bar{b}\gamma\gamma$ following the SM-like Higgs boson pairs, the $b\bar{b}\gamma\gamma$ channel excludes the most space of $\kappa_\lambda=\lambda_{HHH}/\lambda_{HHH}^{\rm SM}$.
We thus analyze this most powerful signal of $b\bar{b}\gamma\gamma$ below for the light Higgs pair as well.
The major SM backgrounds are from the $t\bar{t}H$ and $ZH$ followed by $H\to \gamma\gamma$~\cite{Aaboud:2018ftw}.
We embed the Higgs pair code written in Ref.~\cite{Hespel:2014sla} into MadGraph5\_aMC@NLO~\cite{Alwall:2014hca} to generate signal and background events.
The hadronization and the parton showering are performed by Pythia 8~\cite{Sjostrand:2014zea}.
To simulate the detector effects, we adopt Delphes-3.4.2~\cite{deFavereau:2013fsa} released for detector simulation and event reconstruction by including the beta card for HL-LHC study.
We select some benchmark points passing the above constraints in both Type-I and Type-II 2HDM for the following analysis, as shown in Tab.~\ref{Benchmark}.

\begin{table}[tb]
\begin{center}
\resizebox{\textwidth}{!}{
\begin{tabular}{|c|c|c|c|c|c|c|c|}
\hline
Benchmark & $M_h$ (GeV) & $M_A$ (GeV)  & $\tan\beta$ & $m_{12}^2$ (GeV$^2$) & BR($h\to b\bar{b}$) & BR($h\to \gamma\gamma$) & $\sigma(hh)$ (fb)
\\ \hline
Type I-A & 70 & 314 & 0.29 & 622.8 & 0.85 & $5.1\times 10^{-5}$ & 5448.86\\
Type I-B & 80 & 487.5 & 0.33 & 1334 & 0.845 & $6.9\times 10^{-5}$ & 3283.95\\
Type I-C & 90 & 235.5 & 0.31 & 2004.5 & 0.835 & $9.05\times 10^{-5}$ & 4314.51\\
Type I-D & 100 & 445.9 & 0.55 & 2999 & 0.82 & $1.14\times 10^{-4}$ & 355.373\\
\hline
Type II-A & 70 & 495 & 0.3 & 419.6 & 0.074 & $6.79\times 10^{-4}$ & 4450.96\\
Type II-B & 80 & 268.9 & 0.69 & 6354 & 0.63 & $2.64\times 10^{-4}$ & 263.307\\
Type II-C & 90 & 563 & 0.58 & 4969.6 & 0.456 & $4.98\times 10^{-4}$ & 415.445\\
Type II-D & 100 & 527.9 & 0.84 & 8316.4 & 0.72 & $2.3\times 10^{-4}$ & 90.5083\\
\hline
\end{tabular}
}
\end{center}
\caption{
Benchmark points with $\sin(\beta-\alpha)=0$ for the LHC analysis.
}
\label{Benchmark}
\end{table}

\begin{figure}[htb]
\begin{center}
\includegraphics[width=0.45\textwidth]{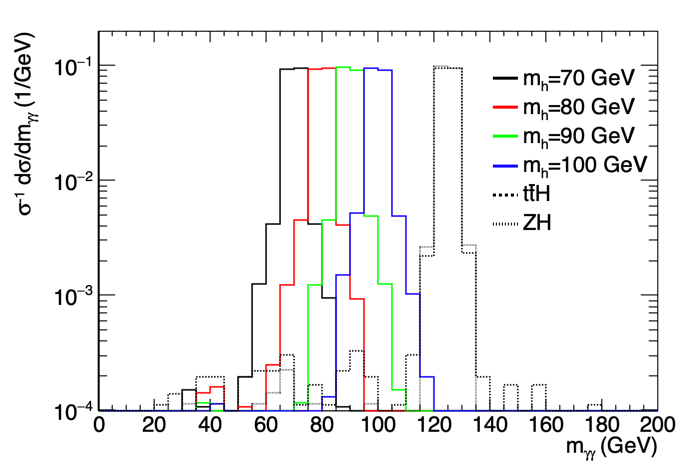}
\includegraphics[width=0.45\textwidth]{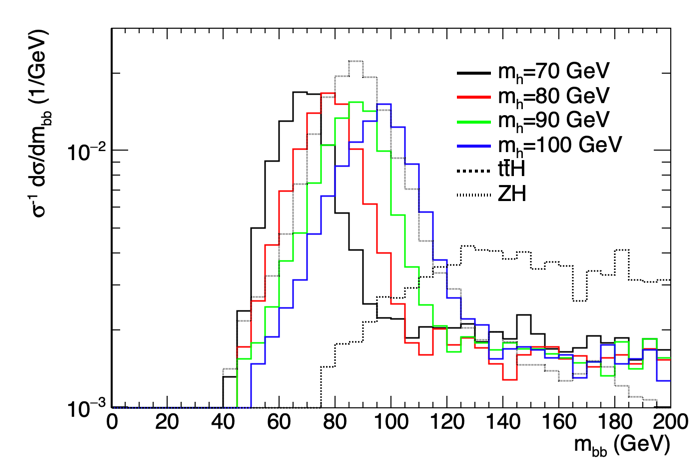}
\end{center}
\caption{The differential distributions for $m_{\gamma\gamma}$ and $m_{b\bar{b}}$ after the basic cuts.}
\label{fig:pphhdis}
\end{figure}

\begin{figure}[!tbp]
\begin{center}
\includegraphics[width=0.9\textwidth]{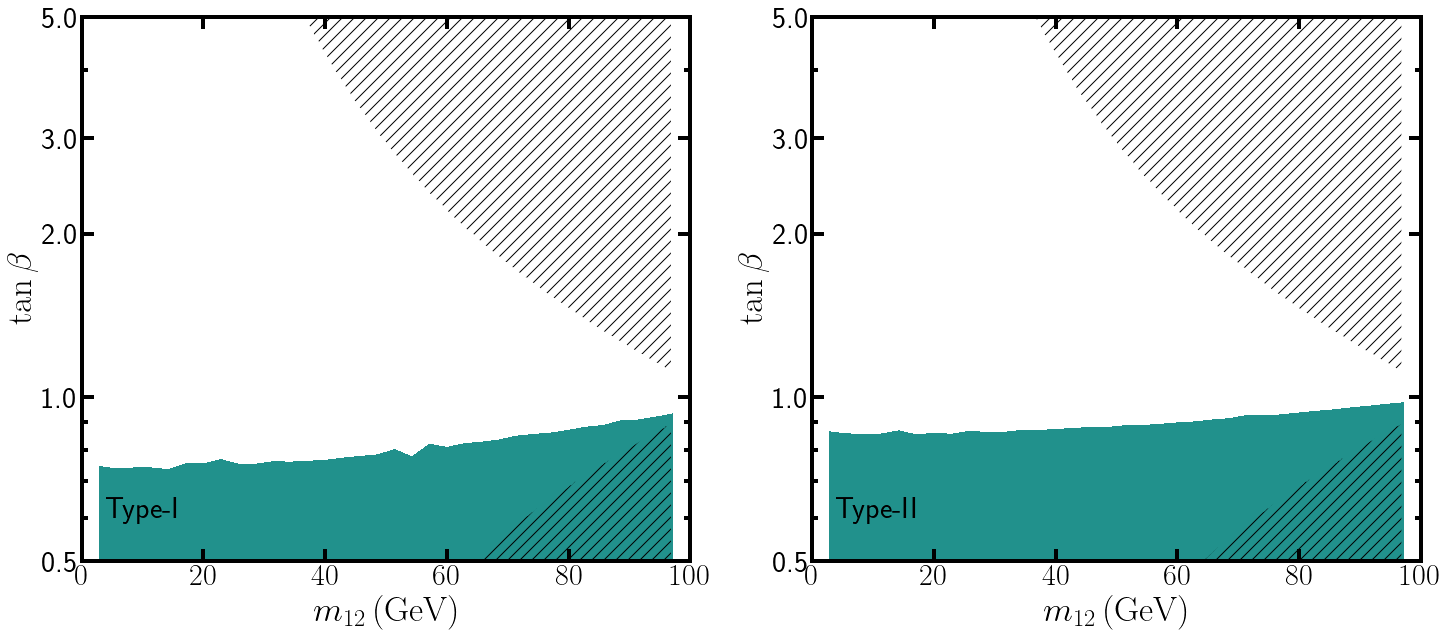}
\end{center}
\caption{
The significances $S/\sqrt{B}$ at the LHC $14\,\TeV$ run with the integrated luminosity of $3~{\rm ab}^{-1}$, in the plane of $m_{12}$ vs. $\tan\beta$ for $M_h=80$ GeV case.
The green region denotes the significance $S/\sqrt{B}>5$.
The perturbative unitarity and stability constraints are shown in the grey shaded regions.
}
\label{fig:SsqrtB}
\end{figure}

We now utilize the $b\bar{b}\gamma\gamma$ channel to search for the light Higgs boson pair productions.
Two leading photons are required to satisfy~\cite{Aaboud:2018ftw,Aad:2019uzh}
\begin{eqnarray}
E_T^{\rm 1st(2nd)}/m_{\gamma\gamma} > 0.35~(0.25)\;, \quad |\eta_\gamma|<2.37\;,
\end{eqnarray}
with $m_{\gamma\gamma}$ being the invariant mass of the diphoton.
For two high-$p_T$ isolated jets, we require at least one $b$-jet with $b$-tagging efficiency of $70\,\%$ and the basic requirements of~\cite{Aaboud:2018ftw,Aad:2019uzh}
\begin{eqnarray}
p_T^{\rm 1st(2nd)}>40~(15)~{\rm GeV}\;, \quad |\eta_j|<2.5\;.
\end{eqnarray}
The two $b$-jets and di-photons are further required to be isolated, i.e. $\Delta R_{bb}$, $\Delta R_{\gamma\gamma}$ and $\Delta R_{b\gamma}>0.4$.
The differential distributions for the invariant masses of $m_{\gamma\gamma}$ and $m_{b\bar{b}}$ after the above basic cuts are displayed in Fig.~\ref{fig:pphhdis}, and one can see clear resonance peaks in the distributions of invariant masses.
Next, we select the events satisfying the invariant mass window
\begin{eqnarray}
|m_{\gamma\gamma}-M_h|<10~{\rm GeV}\;.
\end{eqnarray}
One can see that this cut reduces the backgrounds significantly from the cut efficiencies shown in Tab.~\ref{Selection}.
The significances of $S/\sqrt{B}$ with the integrated luminosity of 300 fb$^{-1}$ or 3 ab$^{-1}$ and the needed luminosity for $5\sigma$ discovery are also shown.
To discovery our benchmarks, one needs the integrated luminosity to be less than 1 fb$^{-1}$ and about 100 fb$^{-1}$ at most.
Finally, in Fig.~\ref{fig:SsqrtB}, the significances of $S/\sqrt{B}$ at the LHC $14\,\TeV$ run with the luminosity of $3~{\rm ab}^{-1}$ are given in the $(m_{12}\,,\tan\beta)$ plane for the $M_h=80$ GeV case.
The green regions denote the discovery significance $S/\sqrt{B}>5$, while the grey shaded regions represent the theoretical constraints.
It turns out that the most relevant constraint comes from the stability bound to $m_{12}$, as shown in Eq.~\eqref{eq:m12_upper}.

\begin{table}[tb]
\begin{center}
\begin{tabular}{|c|c|c|c|c|}
\hline
Benchmark & $m_h=70$ GeV & 80 GeV & 90 GeV & 100 GeV \\ \hline
$\sigma\times {\rm BR}$ (fb) & &  & & \\ % 0.187 \\
Type-I, Type-II & 0.47, 0.45 & 0.38, 0.088 & 0.65, 0.19 & 0.066, 0.03 \\ % 0.7 \\
$t\bar{t}H$, $ZH$ & 0.32, 0.14 & 0.32, 0.14 & 0.32, 0.14 & 0.32, 0.14 \\ \hline
$\sigma\times {\rm BR}\times \epsilon$ (fb) & &  & & \\
Type-I, Type-II & 0.087, 0.083 & 0.086, 0.02 & 0.16, 0.045 & 0.016, 0.0074\\
$t\bar{t}H$, $ZH$ & 0.046, 0.02 & 0.046, 0.02 & 0.046, 0.02 & 0.046, 0.02 \\ \hline
$\sigma\times {\rm BR}\times \epsilon'$ (fb) & &  & & \\
Type-I, Type-II & 0.086, 0.081 & 0.085, 0.019 & 0.15, 0.044 & 0.016, 0.0072\\
$t\bar{t}H(\times 10^{-4})$, $ZH(\times 10^{-4})$ & 1.9, 0.53 & 1.4, 0.22  & 2.0, 0.22  & 1.7, 0.31  \\
\hline
$S/\sqrt{B}$ & &  & & \\
Type-I, Type-II ($\mL=$300 fb$^{-1}$) & 96, 90 & 116, 26 & 174, 51  & 19.5, 8.8 \\
Type-I, Type-II ($\mL=$3 ab$^{-1}$) & 302, 285 & 366, 82 & 551, 162  & 61.8, 27.8\\
\hline
$\mL$ (fb$^{-1}$) & &  & & \\
Type-I, Type-II (5$\sigma$) & 0.82, 0.96 & 0.56, 11 & 0.25, 2.9 & 19.6, 97\\
\hline
\end{tabular}
\end{center}
\caption{
The production cross section of signals and backgrounds times decay branching ratios and cut efficiency for the considered benchmark points.
Here $\epsilon$ and $\epsilon'$ denote the efficiency after basic cuts only and basic+$|m_{\gamma\gamma}- M_h|<10$ GeV cuts, respectively.
The significance $S/\sqrt{B}$ and the needed luminosity for $5\sigma$ significance are also shown.
}
\label{Selection}
\end{table}

%###################################################################
%%%%%%%%%%%%%%%%%%%%%%%%%%%%%%%%
\section{Conclusion}
\label{section:conclusion}
%%%%%%%%%%%%%%%%%%%%%%%%%%%%%%%%

In this work, we have studied the pair production of the light Higgs boson at the LHC.
For illustration We take the 2HDM where the heavier neutral CP-even Higgs $H$ is the observed 125 GeV SM-like Higgs boson and there is a lighter Higgs boson $h$.
This scenario exists in the alignment limit of $\sin(\beta-\alpha)\simeq 0$. We take into account the theoretical constraints and those from the SM-like Higgs exotic decay, the 125 GeV Higgs global fit and the direct LHC searches. It turns out that the global signal fit to the heavy $H$ as the 125 GeV Higgs boson places stringent constraints on $m_{12}$ and the trilinear Higgs coupling $\lambda_{Hhh}$ for $M_H>2M_h$ case. Only a small regime around $\tan\beta\simeq 1$ is survived. For $M_H<2M_h$, relatively small $\tan\beta$ region is excluded by the $h\to \gamma\gamma$ search as the Yukawa coupling to the top quark is enhanced.

We thus focus on the $M_h>M_H/2$ case and take the most powerful signal of $b\bar{b}\gamma\gamma$ following the $h$ pairs in the analysis of light Higgs pair production at collider. To discover the benchmark points passing the above constraints, we find the needed luminosity can be less than 1 fb$^{-1}$ and about 100 fb$^{-1}$ at most at the 14 TeV LHC. The future high-luminosity LHC can probe this light Higgs scenario in the region of  $\tan\beta\lesssim 1$ and $0<m_{12}<100$ GeV allowed by the theoretical constraints.

%###################################################################

\section*{Acknowledgments}

NC is partially supported by the National Natural Science Foundation of China (under Grant No. 11575176).
TL is supported by the National Natural Science Foundation of China (Grant No. 11975129) and ``the Fundamental Research Funds for the Central Universities'', Nankai University (Grant No. 63196013).
The work of NC and TL is also supported in part by the National Natural Science Foundation of China (Grant No. 12035008).
WS is supported by the Australian Research Council (ARC) Centre of Excellence for Dark Matter Particle Physics (CE200100008).
YW is supported by the Natural Sciences and Engineering Research Council of Canada (NSERC).

%\newpage

%###################################################################

\bibliography{references}

\end{document}